\documentclass[oneside,english,usenatbib]{mn2e}
\usepackage[T1]{fontenc}
\usepackage[latin9]{inputenc}
\usepackage[fleqn]{amsmath}
\usepackage[pdftex]{graphicx,color}
\usepackage{soul}
\usepackage{subfigure}
\usepackage{mathptmx}
\usepackage{babel}

\DeclareMathOperator{\erfc}{erfc}

\begin{document}
\global\long\def\ud{\rmn{d}}
\global\long\def\loc{x_{\rmn{loc}}}

\title[Non-virialised clusters for DE--DM interaction]{Non-virialised clusters for detection of Dark Energy--Dark Matter interaction}

\author[M. Le~Delliou et al.]{M. Le~Delliou$^{1,2}$\thanks{Email: delliou@ift.unesp.br (MLeD); mrafael@if.usp.br (RJFM); gastao@astro.iag.usp.br (GBLN); eabdalla@if.usp.br (EA)}, R. J. F. Marcondes$^{2}$\footnotemark[1], G. B. Lima Neto$^{3}$\footnotemark[1] and E. Abdalla$^{2}$\footnotemark[1]\\
$^{1}$Instituto de F\'isica Te\'orica -- Universidade Estadual Paulista, R. Dr. Bento Teobaldo Ferraz, 271 - Bloco II, 
Barra-Funda \\ CEP 01140-070 - S\~ao Paulo, SP, Brazil\\
$^{2}$Departamento de F\'isica Matem\'atica -- Instituto de F\'isica -- Universidade de S\~ao Paulo, Rua do Mat\~ao Travessa R Nr.187 \\ CEP 05508-090, S\~ao Paulo/SP, Brazil \\
$^{3}$Instituto de Astronomia, Geof\'isica e Ci\^encias Atmosf\'ericas, Universidade de S\~ao Paulo, Rua do Mat\~ao 1226, Cidade Universit\'aria, \\ CEP 05508-090, S\~ao Paulo/SP, Brazil}

\date{Send offprint requests to: MLeD\\
Accepted ... Received ...; in original form ...}
\maketitle

\begin{abstract}
    The observation of galaxy and gas distributions, as well as cosmological simulations in a $\Lambda$CDM Universe, suggests that clusters of galaxies are still accreting mass and are not expected to be in equilibrium.
In this work, we investigate the possibility to evaluate the departure from virial equilibrium in order to detect, in that balance, effects from a Dark Matter--Dark Energy interaction. We continue, from previous works, using a simple model of interacting dark sector, the Layzer--Irvine equation for dynamical virial evolution, and employ optical observations in order to obtain the mass profiles through weak lensing and X-ray observations giving the intracluster gas temperatures. Through a Monte Carlo method, we generate, for a set of clusters, measurements of observed virial ratios, interaction strength, rest virial ratio and departure from equilibrium factors. We found a compounded interaction strength of $-1.99^{+2.56}_{-16.00}$, compatible with no interaction, but also a compounded rest virial ratio of $-0.79 \pm 0.13$, which would entail a $2\sigma$ detection. We confirm quantitatively that clusters of galaxies are out of equilibrium but further investigation is needed to constrain a possible interaction in the dark sector.
\end{abstract}


\pagerange{\pageref{firstpage}--\pageref{lastpage}}

\pubyear{2014}

\label{firstpage}

\section{Introduction}

The surprising result of cosmological accelerated expansion, discovered
at first from supernovae data \citep{Perlmutter:1997zf}, and the
long sought Dark Matter seen around clumped matter \citep{Zwicky:1933gu,Zwicky:1937zza},
found their importance combined and confirmed by the results of Cosmological Microwave Background
observations \citep{Planck2013Cosm}. This leads to the fair confidence on the existence of
a dark sector, composed by an inert Cold Dark Matter accounting for
$\sim 27\%$ of the Universe and a rather mysterious Dark Energy, accelerating
the cosmos, accounting for another $\sim 68\%$ \citep{Planck2013Cosm}.

The nature of these components is largely unknown. Several candidates
for Dark Matter appeared in the literature, with no preference from
observations. For Dark Energy, a cosmological constant (de Sitter
space type) is the usual observationally preferred choice and forms
the basis of the so-called standard, or concordance, cosmology \citep{Hinshaw:2012aka,Bennett:2012zja,Ade:2012vda,Ade:2013ktc}.

However, such cosmological constant is an awful choice from the theoretical point
of view: it has been argued that it differs by 120 orders of magnitude
from a reasonable field theory estimation \citep{Weinberg:1988cp,Weinberg:2008zzc}.
From observations, the fact that its energy density is of the same
order of magnitude today as the Dark Matter is a mystery: both depend
on cosmological time in a completely different way, leading to the
so-called coincidence problem \citep{Amendola:1999er,TocchiniValentini:2001ty,Zimdahl:2001ar,Zimdahl:2002zb}.

From the observational standpoint it turns out to be very difficult
to distinguish a clear departure from the standard cosmological model.
Indeed, cosmology is reasonably well described by a positive cosmological constant in a framework of a set of Einstein
equations. A redshift dependent equation of state of Dark Energy is
possible but the error bars obtained from observations so far are still consistent
(to one or two standard deviations) with the standard cosmological
model, although such model is theoretically unsatisfactory as discussed above.

On the contrary, dynamic Dark Energy was theoretically introduced in attempts to solve the fine tuning and coincidence problems described above \citep{Caldwell:1997ii,Copeland:1997et,Zlatev:1998tr}, first using quintessence models such as \citep{Wetterich:1988,Ratra:1988}, then opening to other models, such as $k$-essence \citep{ArmendarizPicon:2000ah} or the generalised Chaplygin gas \citep{Bilic:2001cg,Bento:2002ps}.
In this framework, it was pointed out that, since the main components of the universe in the standard cosmological models are yet unknown, that dark sector interactions, that is the non-minimal coupling between Dark Matter (DM) and Dark Energy (DE), would be the most natural model to consider \citep{Amendola:1999er}. Such models are not ruled out by observations \citep{Pettorino:2012ts,Pettorino:2013oxa}.

One possibility that has been put forward a few years ago is that
interaction can spoil the virial theorem (or else the Layzer--Irvine
equation in General Relativity) \citep{Bertolami:2007zm,LeDelliou:2007am,Bertolami:2008rz,Bertolami:2007tq,Bertolami:2012yp}.
Such a possibility has been checked by different groups \citep{Abdalla:2007rd,Abdalla:2009mt,He:2010ta}
and the results are still not completely settled, although they all point towards the existence of an interacting dark sector. 
Indeed,
using a series of data from clusters, we can infer, under some not
too severe conditions, whether DE can play the role of a
hidden external driving force by means of the dark sector interaction
and thus apparently spoil the virial theorem leaving a trace of the
would-be interaction. In such a case, some limits on the interaction
can be obtained.

In this paper we consider the virial condition and investigate if
it can be clearly imposed observationally, by enquiring into
cluster data and their corresponding observations in the optical and X-ray bands. These bands were chosen because, from deep optical imaging it is possible to derive the total mass distribution through weak-lensing effect of background galaxies, and from X-ray observations we can determine the density and temperature of the intracluster gas, which traces the cluster gravitational potential well.

In previous papers, deviations from the virial theorem,
as described in the Layzer--Irvine equation, were searched in very
relaxed clusters \citep{Bertolami:2007zm,LeDelliou:2007am,Bertolami:2008rz,Bertolami:2007tq,Bertolami:2012yp}
or in large sets of clusters \citep{Abdalla:2007rd,Abdalla:2009mt,He:2010ta}.
In \citet{Bertolami:2007zm}, optical (weak lensing) and X-rays  data
were used to provide the first hint of a detection of a putative violation
of the Equivalence Principle in the virial state of 
Abell 586. Subsequent works aimed at refining the method on Abell 586 \citep{LeDelliou:2007am,Bertolami:2008rz,Bertolami:2007tq}
and introduced another relaxed cluster candidate \citep[Abell 1689,][]{Bertolami:2012yp}.
In \citet{Abdalla:2007rd}, optical and X-ray data were
compared and a hint of systematic bias in the estimation of masses,
if the usual virial conditions are employed, was found. In a subsequent
paper \citep{Abdalla:2009mt}, the work was extended to a larger set
of clusters. In both cases, it was argued that such signals might
point to new physics to be uncovered and more specifically to a correction
of the virial theorem due to a would-be interaction of Dark Matter
with a hidden sector, i.e. Dark Energy, driving the system away from
the virial equilibrium.

In the present work, we address some of these issues raised previously,
digging into the internal structure of the clusters, in order to find,
in a very direct way, not only the kinetic versus potential energy
--- the virial ratio --- from each system (cluster), but also to use
the full dynamics of the Layzer--Irvine equation, including the same fluid model for DM and DE as used in previous works, in order to extract
an assessment of the departure from self-gravitating equilibrium.

The next section describes the techniques
used for extracting relevant dynamical quantities from the data. Sec.~\ref{sec:Computations-of-the} compiles the relevant data and summarises
their treatment. The results obtained are given in Sec.~\ref{sec:Analysis-of-the},
and we conclude with a discussion in Sec.~\ref{sec:Discussion-and-Conclusions}. We adopt whenever necessary the standard flat $\Lambda$CDM parameters: $\Omega_M = 0.32$ for the Matter density parameter, $\Omega_{DE} = 0.68$ for the DE parameter and Hubble constant $H_0 = 67.0~\rmn{km}~\rmn{s}^{-1}~\rmn{Mpc}^{-1}$, according to the best-fit parameters from Planck temperature combined with WMAP polarization at low multipoles \citep{Planck2013Cosm}.

\section{Virial ratio and Departure from Equilibrium for Non-Virialised Clusters}
\label{sec:Virial-ratio-and}

In \citet{Abdalla:2007rd}, a model of interaction involving the coupling
of both DM and DE was introduced for which
the Layzer--Irvine equation was deduced. We adopt a form, in what follows,
that is inspired by such model.

\subsection{The Layzer--Irvine model with interaction}

The Layzer--Irvine equation \citep{Peebles} is a model of dynamical
virial balance where the disturbance is simplified in the expansion
of the background universe.

\subsubsection{The interaction model}
\label{sss:intmodel}

Following \citet{Abdalla:2007rd,Abdalla:2009mt} and \citet{He:2010ta}, we model
the DM--DE interaction via a heat flux between the two species, themselves treated as fluids with constant equation-of-state parameters, in their
partial Bianchi identities in a Friedmann--Lema\^itre--Robertson--Walker (FLRW) background model
\begin{gather}
\dot{\rho}_{DM}+3H\rho_{DM}=3H\left(\xi_{1}\rho_{DM}+\xi_{2}\rho_{DE}\right),\label{eq:DMcons} \\
\dot{\rho}_{DE}+3H\rho_{DE}(1+\omega_{DE})=-3H\left(\xi_{1}\rho_{DM}+\xi_{2}\rho_{DE}\right).\label{eq:DEcons}
\end{gather}
Here we have used the indices to refer to species, the couplings $\xi_{1}$
and $\xi_{2}$, hereafter interaction strengths, the densities denoted by $\rho_{x}$ while their corresponding
equation of state reads $\omega_{x}$ and the FLRW Hubble parameter
$H$. In what follows we shall simplify the interaction by choosing
$\xi_{1}=\frac{\xi}{18}$ and $\xi_{2}=-\frac{\xi}{6}\frac{\rho_{DM}}{\rho_{DE}}$
and writing $\rho_{DM}=\rho$, yielding a positive flux $\frac{\xi}{3}H\rho$
towards DE when all terms are positive.

The resulting Layzer--Irvine equation for the DM component \citep{He:2010ta},
referring to its kinetic energy density with $\rho_{K}$ while its
potential energy density writes $\rho_{W}$, reads
\begin{align}
\dot{\rho}+H\left[\left(2-\frac{\xi}{3}\right)\rho_{K}+\left(1-\frac{\xi}{3}\right)\rho_{W}\right] & =0.\label{eq:L-I}
\end{align}
In case of equilibrium, the time derivative vanishes and yield the
interacting virial balance as 
\begin{align}
\frac{\rho_{K}}{\rho_{W}} &= -\frac{1-\frac{\xi}{3}}{2-\frac{\xi}{3}}.\label{eq:KsW}
\end{align}
However, in this work, we want to take into account departures from
equilibrium.

Note that certainty of convergence of the energy density towards equilibrium, together with other magnitude restriction considerations \citep[e.g.][]{He:2011}, prescribes from Eq.~(\ref{eq:L-I}) to exclude values of $\xi$ higher than 3.

\subsubsection{The Non-Virialised model}
\label{sss:nvmodel}

In order to simplify the calculation, we approximate the departure
of $\rho_{K}$ from balance as proportional to the departure of $\rho_{W}$
and introduce Eq. (\ref{eq:KsW}) into (\ref{eq:L-I}) to get 
\begin{align}
    \label{eq:nvmodel}
\left[1-\frac{1-\frac{\xi}{3}}{2-\frac{\xi}{3}}\right]\dot{\rho}_{W} &= -H\left[\left(2-\frac{\xi}{3}\right)\rho_{K}+\left(1-\frac{\xi}{3}\right)\rho_{W}\right]
\end{align}
 so 
\begin{align}
\frac{\rho_{K}}{\rho_{W}} &= -\frac{1-\frac{\xi}{3}}{2-\frac{\xi}{3}}-\frac{1}{\left(2-\frac{\xi}{3}\right)^{2}}\frac{\dot{\rho}_{W}}{H\rho_{W}}.\label{eq:Virial+Out}
\end{align}
 Here we find that the true virial balance (\ref{eq:KsW}) is corrected
by a term we call departure from equilibrium (DfE). For our approximation
(\ref{eq:KsW}) to remain valid, the DM halo has to be close to virial
balance, so we have to check that 
\begin{align}
    \left| \frac{\dot{\rho}_{W}}{H\rho_{W}} \right| &\ll  \left(2-\frac{\xi}{3}\right)\left(1-\frac{\xi}{3}\right).\label{eq:Close2EqCond}
\end{align}
 In that case, $\rho_{K}$, $\rho_{W}$, $\dot{\rho}_{W}$ and $H$
being observable, we can solve Eq. (\ref{eq:Virial+Out}) for the
coupling and obtain 
\begin{align}
\xi & =3\left(\frac{3+4\frac{\rho_{K}}{\rho_{W}}-\sqrt{1-\frac{4\dot{\rho}_{W}}{H\rho_{W}}\left(1+\frac{\rho_{K}}{\rho_{W}}\right)}}{2\left(1+\frac{\rho_{K}}{\rho_{W}}\right)}\right).\label{eq:Coupling}
\end{align}
The expression verifies the classical, non-interacting and virialised,
result: $\xi=0$ for $\dot{\rho}_{W}=0$ and $\frac{\rho_{K}}{\rho_{W}}=-\frac{1}{2}$.
Note that this equation is singular at $\frac{\rho_K}{\rho_W} = -1$, which originates in the left-hand side of Eq.~(\ref{eq:nvmodel}) and corresponds to infinite $\xi$.

Note that this method sums all possible dynamical sources of deviation from the stationary true virial balance into this DfE factor, including sources such as simulations imperfectly stationary states caused by finite integration time and/or numerical noise and, for observed clusters, the possibility of recent merger events kinetic remnants, inflows from neighbouring clusters or baryonic sources of kinetic injection in the cluster (supernova feedback, baryons-DM dynamical friction) \citep{Kravtsov2012}.
The evaluation of any sources of deviations of the real virial balance from the Virial Theorem classical value \citep{BinneyandTremaine} are encoded in our simple model of DM--DE interaction as the presently conceived source. Thus far, no other source of deviation claims to have an effect that would not vanish at real equilibrium. We are henceforth justified in considering this evaluation as a handle on possible interaction and its subsequent violation of the Equivalence Principle between baryons and DM. We note however that given the simplifications, this work should be considered as a proof of concept for such an evaluation on unbalanced clusters.

We now need to evaluate $\rho_{K}$, $\rho_{W}$, $\dot{\rho}_{W}$
and $H$ from cluster observations.
However, as we will see in Sec.~\ref{sss:VandDeval}, the factor $\frac{\dot\rho_W}{H\rho_W}$ in the DfE is not a pure observable and depends on $\xi$.

\subsection{The evaluation from clusters}
\label{sec:evalclusters}
As will be seen in Sec.~3, for each cluster we essentially have access to the total mass distribution through weak lensing observations given in the form of Navarro, Frenk and White (NFW) profile parameter fits, and hence can derive its potential energy, and to the evaluation of the cluster's kinetic state through its X-ray temperature, that we can then transform into an evaluation of its kinetic energy. In what follows, we provide the framework to make contact between such observables and the theoretical scheme we have presented above.

\subsubsection{The NFW density and Weak Lensing mass}

The NFW density profile \citep{Navarro:1995iw} found in $N$-body simulations
is commonly used to fit observed clusters in order to parametrize their mass distribution. The classical
form of the NFW profile involves the radius of logarithmic slope change
$r_{0}$ and the corresponding density $\rho_{0}$. Assuming the virial
radius of clusters lies at a density contrast of about 200 above the
background density for clusters, the NFW concentration parameter can
be defined as $c=\frac{r_{200}}{r_{0}}$. Moreover, the profile's
integration yields a mass profile from which $M_{200}$ can be extracted.
In terms of $c$, $r_{200}$ and $M_{200}$, the mass and density
profiles read
\begin{align}
    \label{eq:Massprofile}
M & =\frac{M_{200}}{\ln\left(1+c\right)-\frac{c}{1+c}}\left[\ln\left(1+c\frac{r}{r_{200}}\right)-\frac{c\frac{r}{r_{200}}}{1+c\frac{r}{r_{200}}}\right],\\
\rho & =\frac{M_{200}}{4\pi r_{200}^{3}\left[\ln\left(1+c\right)-\frac{c}{1+c}\right]}\frac{c^{2}}{\frac{r}{r_{200}}\left(1+c\frac{r}{r_{200}}\right)^{2}}.
\end{align}
The parameter describing the best-fit NFW profile, $c$ and $r_{200}$ (or $M_{200}$), can be obtained from the mass reconstruction technique, which requires deep imaging with small point-spread-function (PSF) that can detect the weak lensing effect.

\subsubsection{The potential and kinetic energy density evaluations}

With the previously presented NFW mass profile, the density of potential
energy is straightforwardly integrated
\begin{align}
    \rho_{W} & \equiv-\frac{4\pi}{\frac{4}{3}\pi r_{200}^{3}}\int_{0}^{r_{200}}\frac{\rho(r)GM(r)}{r}r^{2}\, \ud r  =-\frac{3GM_{200}^{2}}{4\pi r_{200}^{4}f_{c}}, \label{eq:PotEnDen}
 \end{align}
with
\begin{align}
    f_{c} & \equiv\frac{(1+c)\left[\ln(1+c)-c(1+c)^{-1}\right]^{2}}{c\left[\frac{1}{2}\left\lbrace (1+c)-(1+c)^{-1}\right\rbrace -\ln\left(1+c\right)\right]}.
\end{align}
On the other hand, in order to evaluate the kinetic state of the cluster, we use published 
X-rays observations, where we just need to obtain the X-ray temperature to get the equipartition
formula
\begin{align}
    \rho_{K} & =\frac{3}{2}N\frac{k_{\rmn{B}} T_{X}}{V}=\frac{9}{8\pi}\frac{M_{200}}{r_{200}^{3}}\frac{k_{\rmn{B}}T_{X}}{\mu m_{\rmn{H}}},
\end{align}
gauging the equivalent number of particles from the total mass given
by weak lensing $\frac{M_{200}}{\mu m_{\rmn{H}}}$, where $\mu$ is the
mean molecular mass in the intracluster gas and $m_{\rmn{H}}$ is the mass of a proton.
The advantage of this method, compared, say, to using the galaxy velocity
dispersion or a scaling relation $\sigma_X$--$T_X$, as in \citet{Bertolami:2007zm}, \citet{LeDelliou:2007am} and \citet{Bertolami:2008rz,Bertolami:2007tq,Bertolami:2012yp},
is that we avoid the error from the scatter in the scaling relation.
We are justified in evaluating the kinetic state of clusters through the single compounded temperature $T_X$, extracted from the X-ray flux of the whole $r_{500}$ central region, as it already largely encompasses the turnaround of the temperature profile \citep{Vik:2005,Pratt:2007,Moretti:2011}, and therefore represents well the total density averaged temperature (the so-called virial temperature).

At this point the virial ratio can be evaluated as
\begin{align}
    \label{eq:Virialratio}
    \frac{\rho_{K}}{\rho_{W}} &= -\frac{3}{2}\frac{r_{200}}{GM_{200}}\frac{k_{\rmn{B}}T_{X}}{\mu m_{\rmn{H}}}f_{c}.
\end{align}

\subsubsection{The virial and departure from equilibrium evaluation}
\label{sss:VandDeval}

The DfE factor can be rewritten as
\begin{align}
-\frac{1}{\left(2-\frac{\xi}{3}\right)^{2}}\frac{\dot{\rho}_{W}}{H\rho_{W}} &= -\frac{1}{\left(2-\frac{\xi}{3}\right)^{2}}\frac{\rho_{W}^{\prime}}{H\rho_{W}}\dot{r}_{200},
\end{align}
where we get from Eq.~(\ref{eq:PotEnDen}) 
\begin{align}
\frac{\mathrm{d}\ln\left(-\rho_{W}\right)}{\mathrm{d}r_{200}}  =\frac{\rho_{W}^{\prime}}{\rho_{W}} = \frac{cg_{c}-3}{r_{200}},
\end{align}
with
\begin{align}
    g_{c} \equiv\frac{\ln\left(1+c\right)-c\left(1+c\right)^{-1}}{\frac{c}{2}\left(c+2\right)-\left(1+c\right)\ln\left(1+c\right)}.
\end{align}
Now remains to evaluate $\dot{r}_{200}$. We can define the kinetic
density using a one-dimensional velocity dispersion, thus defining
$\sigma_{X}^{2}$:
\begin{align}
    \rho_{K} &= \frac{3}{2}\frac{M_{200}}{V}\sigma_{X}^{2}\Rightarrow\sigma_{X}^{2}=\frac{k_{\rmn{B}}T_{X}}{\mu m_{\rmn{H}}}.
\end{align}
We now define the theoretical average velocity dispersion the cluster would have
if it were at virial equilibrium, adiabatically evolving from the
current state (meaning keeping potential energy about constant)
\begin{align}
    \label{eq:thVR}
    \left(\frac{\rho_K}{\rho_{W}}\right)_{th} & =-\frac{1-\frac{\xi}{3}}{2-\frac{\xi}{3}}\text{ with }\rho_{K\, th}=\frac{3}{2}\frac{M_{200}}{V}v_{th}^{2}\\
\Leftrightarrow v_{th}^{2} & =\frac{1}{3}\frac{6-2\xi}{6-\xi}\frac{GM_{200}}{f_{c}r_{200}}.
\end{align}
We can finally evaluate the time evolution of $r_{200}$ by taking
its difference with the velocity dispersion
\begin{align}
    \dot{r}_{200} &= \sqrt{\sigma_{X}^{2}}-\sqrt{v_{th}^{2}}=\sigma_{X}-v_{th},
\end{align}
we obtain
\begin{align}
    -\frac{\dot{\rho}_{W}}{H\left(2-\frac{\xi}{3}\right)^2\rho_{W}} &= -\frac{1}{H\left(2-\frac{\xi}{3}\right)^{2}}\frac{\left(cg_{c}-3\right)}{r_{200}} \times {} \nonumber \\
    \label{eq:Out(xi)}
    & \quad  \times \left(\sqrt{\frac{k_{\rmn{B}}T_{X}}{\mu m_{\rmn{H}}}} - \sqrt{\frac{1}{3}\frac{6-2\xi}{6-\xi}\frac{GM_{200}}{f_{c}r_{200}}}\right).
\end{align}
With this equation, we estimate the departure from equilibrium due to ``standard'' dynamical sources (e.g.~cluster collisions, AGN and supernova feedback, dynamical friction) combining observations and the dark energy model, leaving no room for degeneracy in the determination of $\xi$.
The DfE presented here appears model dependent in its explicit reference to the interaction strength;
however, the method can use any model we want that gives a definite
shift to the virial balance\footnote{
    \begin{align}
        \frac{\rmn{VR} \, \dot{\rho}_{W}}{H c_W\left(1+\rmn{VR}\right)\rho_{W}} &= -\frac{\rmn{VR}}{H c_W\left(1+\rmn{VR}\right)}\frac{\left(cg_{c}-3\right)}{r_{200}} \times {} \nonumber \\
        \label{eq:OutGeneral}
        & \quad  \times \left(\sqrt{\frac{k_{\rmn{B}}T_{X}}{\mu m_{\rmn{H}}}} - \sqrt{-\frac{2}{3}\rmn{VR}\frac{GM_{200}}{f_{c}r_{200}}}\right),
    \end{align} where VR is the virial ratio and $c_W$ is the potential coefficient in the Layzer--Irvine equation.}. In this case, decrease to negative values of $\xi$ from 0, at fixed observations, would decrease the absolute value of clusters with kinetic velocity above their virial one and up to increase those of the opposite case.

\section{Computations of the virial ratios and departure from equilibrium
for a set of Non-Virialised Clusters}
\label{sec:Computations-of-the}
Cosmologically interesting observations of clusters are produced in many surveys and studies such as \citet{Okabe25062010} and \citet{Ade:2012vda}.
In order to maximise our sample, while being able to
separately evaluate from observations the kinetic and potential energy
states of each cluster, we have restricted inputs to weak lensing NFW fit parameters \citep{Navarro:1995iw}, X-ray derived $c_{500}$ NFW fits, and X-ray temperatures.

\subsection{The sample}
In order to try to minimise any systematics due to observational uncertainties, the clusters in our sample should present well determined X-ray gas temperature, as well as NFW profile fitted to the mass distribution obtained with weak lensing observations.

    Most of the 22 clusters in our sample come from \citet{Okabe25062010}. Their NFW profiles are described by best-fit virial masses $M_{\rmn{vir}}$, concentration parameters $c_{\rmn{vir}}$ and masses $M_{200}$ estimated from this three-dimensional model fitting. Those are the Abell clusters A68, A115, A209, A267, A383, A521, A586, A611, A697, A1835, A2219, A2261, A2390, A2631 and also RX J1720.1+2638, RX J2129.6+0005, ZwCl 1454.8+2233 and ZwCl 1459.4+4240. Their weak lensing data are shown in Table~\ref{tbl:subset2}. 
\begin{table}
    \caption{Weak lensing masses $M_{200}$, $M_{\rmn{vir}}$ and concentration $c_{\rmn{vir}}$ for the Okabe's clusters. Masses are in units of $h^{-1}10^{14}M_{\odot}$. }
    \label{tbl:subset2}
    \begin{tabular}{@{}lccc@{}}
        \hline
        Cluster & $M_{200} $ & $M_{\rmn{vir}}$ & $c_{\rmn{vir}}$ \\ \hline
        A68 & $4.45^{+1.75}_{-1.35}$ & $5.49^{+2.56}_{-1.81}$ & $4.02^{+3.36}_{-1.82}$ \\
        A115 & $4.24^{+2.60}_{-1.79}$ & $5.36^{+4.08}_{-2.45}$ & $3.69^{+5.03}_{-2.04}$ \\
        A209 & $10.62^{+2.17}_{-1.81}$ & $14.00^{+3.31}_{-2.60}$ & $2.71^{+0.69}_{-0.60}$ \\
        A267 & $3.23^{+0.82}_{-0.69}$ & $3.85^{+1.08}_{-0.88}$ & $6.00^{+2.11}_{-1.58}$ \\
        A383 & $3.11^{+0.88}_{-0.69}$ & $3.62^{+1.15}_{-0.86}$ & $8.87^{+5.22}_{-3.05}$ \\
        A521 & $4.58^{+1.00}_{-0.88}$ & $5.85^{+1.45}_{-1.22}$ & $3.06^{+1.01}_{-0.79}$ \\
        A586 & $6.29^{+2.26}_{-1.69}$ & $7.37^{+2.89}_{-2.08}$ & $8.38^{+3.52}_{-2.52}$ \\
        A611 & $5.47^{+1.31}_{-1.11}$ & $6.65^{+1.75}_{-1.42}$ & $4.23^{+1.77}_{-1.23}$ \\
        A697 & $9.73^{+1.86}_{-1.61}$ & $12.36^{+2.68}_{-2.21}$ & $2.97^{+0.85}_{-0.69}$ \\        
        A1835 & $10.86^{+2.53}_{-2.08}$ & $13.69^{+3.65}_{-2.86}$ & $3.35^{+0.99}_{-0.79}$ \\
        A2219 & $7.75^{+1.89}_{-1.60}$ & $9.11^{+2.54}_{-2.06}$ & $6.88^{+3.42}_{-2.16}$ \\
        A2261 & $7.97^{+1.51}_{-1.31}$ & $9.49^{+2.01}_{-1.69}$ & $6.04^{+1.71}_{-1.31}$ \\
        A2390 & $6.92^{+1.50}_{-1.29}$ & $8.20^{+1.93}_{-1.63}$ & $6.20^{+1.53}_{-1.28}$ \\
        A2631 & $4.54^{+0.89}_{-0.78}$ & $5.24^{+1.15}_{-0.98}$ & $7.84^{+3.54}_{-2.28}$ \\
        RX J1720 & $3.48^{+1.28}_{-0.99}$ & $4.07^{+1.65}_{-1.22}$ & $8.73^{+5.60}_{-3.08}$ \\
        RX J2129 & $5.29^{+1.76}_{-1.38}$ & $6.71^{+2.73}_{-1.96}$ & $3.32^{+2.16}_{-1.34}$ \\
        ZwCl 1454 & $2.80^{+1.39}_{-1.03}$ & $3.45^{+2.02}_{-1.36}$ & $4.01^{+3.44}_{-1.96}$ \\
        ZwCl 1459 & $3.77^{+1.17}_{-0.98}$ & $4.40^{+1.50}_{-1.20}$ & $6.55^{+3.34}_{-2.18}$ \\
        \hline
    \end{tabular}

    \medskip
    Data from \citet{Okabe25062010}.
\end{table}
We also include four more clusters from \citet{Ade:2012vda}: A520, A963, A1914 and A2034 (data in Table~\ref{tbl:subset1}).
\begin{table}
    \caption{Weak lensing masses $M_{500}$ and concentration $c_{500}$ for the Planck Collaboration's clusters. Masses are in units of $10^{14}M_{\odot}$.}
    \label{tbl:subset1}
    \begin{tabular}{@{}lcc@{}}
        \hline
        Cluster & $M_{500}$ & $c_{500}$ \\
        \hline
        A520 & $4.1^{+1.1}_{-1.2}$ & $1.4 \pm 0.6$ \\
        A963 & $4.2^{+0.9}_{-0.7}$ & $1.2 \pm 0.2$ \\
        A1914 & $4.7^{+1.6}_{-1.9}$ & $2.0 \pm 0.2$ \\
        A2034 & $5.1^{+2.1}_{-2.4}$ & $1.8 \pm 0.3$ \\
        \hline
    \end{tabular}

    \medskip
    Data from \citet{Ade:2012vda}.
\end{table}
Weak lensing masses $M_{500}$ and best fitting NFW concentration parameter $c_{500}$ are given instead of $M_{\rmn{vir}}$, $c_{\rmn{vir}}$ and $M_{200}$ for these clusters.
    However, the error bars for $c_{500}$ were estimated from the X-ray data, since they are not given by \citet{Ade:2012vda}.
The spectroscopically determined temperatures $T_{X}$ are measured within $r_{500}$ and are all given in \citet{Ade:2012vda}, with the exceptions of A115 and A697 from \citet{Landry2013} and A611 from \citet{Kenneth2008}. 
Uncertainties correspond to $1\sigma$ confidence level (C.L.). 
Errors in redshifts (see Table~\ref{tab:obsdata}) are not specified but can be safely neglected compared to the errors in other quantities (the typical spectroscopic redshift error is around 1\%).

\subsubsection{Uniforming the NFW profiles}
\label{sss:unif}
To compute the potential energy, we use weak lensing observational fittings of NFW profiles. These fittings, justified by the observers confidence in their reliability \citep{Okabe25062010,Ettori2013}, are further vindicated by interacting DE models from simulations \citep[e. g.][]{Baldi:2008ay,Carlesi:2014faa} which all agree with the NFW shape.

We want to have all NFW profiles parametrised by $M_{200}$ and $c$.
In general, within a radius $r_{\Delta}$, i.e.~the scale for which the density is equal to $\Delta$ times the critical density, we have
\begin{align}
    \Delta = \frac{M_{\Delta}}{\frac{4}{3}\pi r_{\Delta}^3 \rho_c(z)},
\end{align}
which results in an expression for the radius given the mass:
\begin{align}
    \label{eq:rDelta}
    r_{\Delta} = \frac{1}{H(z)} \sqrt[3]{\frac{2 G M_{\Delta} H(z)}{\Delta}}.
\end{align}

For the latter set of clusters, with NFW profiles specified by $M_{500}$ and $c_{500} = \frac{r_{500}}{r_0}$ (rather than $c_{200}$, which we called just $c$), shown in Table~{\ref{tbl:subset1}}, the parameter $r_0 = \frac{r_{200}}{c} =  \frac{r_{500}}{c_{500}}$ comes immediately by using Eq.~(\ref{eq:rDelta}) for $\Delta=500$. Using \citeauthor{sympy2014} in python, we get $M_{200}$ and $r_{200}$ by solving simultaneously an equation similar to (\ref{eq:Massprofile}) (but parametrized by $M_{500}$ and $c_{500}$ and evaluated at $r_{200}$ to give $M_{200}$) and Eq.~(\ref{eq:rDelta}) with $\Delta=200$:
\begin{align}
        M_{200} &= \frac{M_{500}}{\ln(1 + c_{500}) - \frac{c_{500}}{1+ c_{500}}} \left[ \ln \left(1 + c_{500} \tfrac{r_{200}}{r_{500}} \right) - \frac{c_{500} \frac{r_{200}}{r_{500}}}{1 + c_{500} \frac{r_{200}}{r_{500}}} \right], \\
        r_{200} &= \frac{1}{H(z)} \sqrt[3]{\frac{G M_{200} H(z)}{100}}.
\end{align}
Then we can finally compute $c = \frac{r_{200}}{r_0}$. 

For the former set, with NFW profiles specified by $M_{200}$, $M_{\rmn{vir}}$ and $c_{\rmn{vir}}$ (Table~\ref{tbl:subset2}), ``vir'' would correspond to some $\Delta_{\rmn{vir}}$ around 200 but this value can vary with the redshift. Then we proceed as follows. We compute $r_{200}$ from Eq.~(\ref{eq:rDelta}) and solve
\begin{align}
    M_{200} &= \frac{M_{\rmn{vir}}}{\ln(1+c_{\rmn{vir}}) - \frac{c_{\rmn{vir}}}{1+c_{\rmn{vir}}} } \left[ \ln \left(1 + c_{\rmn{vir}} \tfrac{r_{200}}{r_{\rmn{vir}}} \right) - \frac{c_{\rmn{vir}} \frac{r_{200}}{r_{\rmn{vir}}}}{1 + c_{\rmn{vir}} \frac{r_{200}}{r_{\rmn{vir}}}} \right] \label{eq:solrvir}
\end{align}
for $r_{\rmn{vir}}$. $\Delta_{\rmn{vir}}$ can also be determined now with $r_{\rmn{vir}}$ and $M_{\rmn{vir}}$, inverting Eq.~(\ref{eq:rDelta}). 
Finally, we have $r_0 = \frac{r_{\rmn{vir}}}{c_{\rmn{vir}}}$ and $c = \frac{r_{200}}{r_0}$.
The errors are estimated using a Monte Carlos method that we describe in the next section.
With $c$ and $M_{200}$, we can now proceed to the computation of the virial ratios.

\subsection{Monte Carlo estimation of errors}
\label{ss:MonteCarlo2}

We apply a Monte Carlo method to propagate uncertainties through the numerical solutions. We perform multiple realisations of each cluster, with the observables assuming values that are drawn from a distribution that reflects the $1\sigma$ confidence intervals from the original asymmetrical uncertainties. We then carry the computations for all the realisations of each cluster and analyse the final distribution of the quantities of interest to get their error bars.

\subsubsection{Uncertainties in $M_{200}$ and $c_{200}$}
These two NFW parameters are always positive. In order to guarantee that their uncertainties will not lead to negative values in any of the random realisations, we choose the log-logistic distribution, a non-negative probability distribution that has simple analytical forms for its probability density function (PDF) and cumulative distribution function (CDF). They are given, in terms of the parameters $\alpha$ and $\beta$, by
\begin{align}
    f_X(x; \alpha, \beta) = \frac{\frac{\beta}{\alpha} \left( \frac{x}{\alpha} \right)^{\beta-1}}{\left[ 1 + \left(\frac{x}{\alpha} \right)^{\beta}\right]^2}, \quad F_X(x; \alpha, \beta) = \frac{1}{1 + \left(\frac{x}{\alpha}\right)^{-\beta}},
\end{align}
respectively, for a random variable $X$.

Given an observable $X$ with measured value $x^{+\Delta x_1}_{-\Delta x_2}$,
we would like the Monte Carlo generating distribution to match the following criteria:
\begin{enumerate}
    \item \label{i:mode} The maximum probability coincides with the nominal measure;
    \item \label{ii:CDF} The probability of $X$ lying between $x-\Delta x_2$ and $x+\Delta x_1$ is 68\%;
    \item \label{iii:PDF} The PDF has the same value at the two points $x-\Delta x_2$ and $x+\Delta x_1$, so that the interval in condition \ref{ii:CDF} corresponds to the 68\% most probable values, i.e.~$1\sigma$ C.L.
\end{enumerate}
In the case of the log-logistic distribution, these conditions are translated by:
\begin{enumerate}
    \item $\alpha \left(\frac{\beta-1}{\beta+1}\right)^{1/\beta} = x$ (for $\beta > 1$);
    \item $F_X(x+\Delta x_1; \alpha, \beta) - F_X(x-\Delta x_2; \alpha, \beta) = 0.68$;
    \item $f_X(x-\Delta x_2; \alpha, \beta) = f_X(x+\Delta x_1; \alpha, \beta)$.
\end{enumerate}
But these are too many conditions for a distribution that has only two parameters. We choose to relax condition \ref{i:mode} and solve \ref{ii:CDF} and \ref{iii:PDF} for $\alpha$ and $\beta$. In practice our resulting maximum probabilities usually happen to be very close to $x$.

When extracting the $1\sigma$ C.L., we take the opposite direction and get a best-fit log-logistic PDF for the distributions of $M_{200}$ and $c$, now solving \ref{ii:CDF} and \ref{iii:PDF} for $x-\Delta x_2$ and $x+\Delta x_1$. We assign the maximum probability of the distribution to the nominal value $x$.

We have also used the log-normal distribution to check whether our choice of distribution could be biasing our results.
The log-normal PDF and CDF are given by
\begin{align}
    f_X(x; \mu, \sigma) = \frac{e^{-\left( \ln x - \mu \right)^2/2\sigma^2}}{x \sigma \sqrt{2 \pi}}, \quad F_X(x; \mu,\sigma) = \frac{1}{2} \erfc \left[ - \frac{\ln x - \mu}{\sigma \sqrt{2}} \right],
\end{align}
where $\erfc(x)$ is the complementary error function and $\mu$ and $\sigma$ are the Gaussian parameters of the distribution of $\ln X$.

We applied this Monte Carlo procedure for the clusters in our sample. However, the log-normal distribution could not satisfy our requirements \ref{ii:CDF} and \ref{iii:PDF} for all clusters in the first group. Nevertheless, we were able to verify in the other cases, where the log-normal distribution works, that the confidence intervals obtained with the two distributions are very similar, within a few percent of displacement between their extremities. The maximum probability can vary a little more between the two distributions because \ref{i:mode} is not being satisfied, but we are more concerned with the confidence intervals, since we use uniform distributions for $c$, $M_{200}$ and $T_X$ in the evaluation of the virial ratios, interaction strength and departure from equilibrium. We believe, then, that the use of the log-logistic distribution with the requirements that we propose for the estimation of errors for $M_{200}$ and $c$ is a reasonable choice, as it works for all clusters and the results seem not to be biased.

\subsubsection{Virial ratios and interaction strength fittings}
\label{sss:generalfit}
In Table~\ref{tab:obsdata} we summarise the data used for computation of the virial ratios and interaction strengths according to the steps in Sec.~\ref{sec:evalclusters}.
\begin{table}
    \caption{Redshift, temperature and compiled NFW parameters of the 22 galaxy clusters. Temperatures are given in $\rmn{keV}$ and masses in units of $h^{-1}10^{14}M_{\odot}$.}
    \label{tab:obsdata} %
    \begin{tabular}{@{}lcccc@{}}
        \hline  
        Cluster  & $z$  &  $k_{\rmn{B}}T_{\rmn{X}}$ &  $M_{200}$  & $c$\\
        \hline 
        A68 & $0.255$ & $8.3 \pm 0.3$ & $4.45^{+1.75}_{-1.35}$ & $2.49^{+3.12}_{-1.65}$ \\
        A115 & $0.197$ & $8.9^{+0.6}_{-0.7}$ & $4.24^{+2.60}_{-1.79}$ & $1.86^{+3.52}_{-1.48}$ \\
        A209 & $0.206$ & $6.6 \pm 0.2$ & $10.62^{+2.17}_{-1.81}$ & $1.90^{+0.81}_{-0.63}$ \\
        A267 & $0.230$ & $5.6 \pm 0.1$ & $3.23^{+0.82}_{-0.69}$ & $3.95^{+3.00}_{-1.96}$ \\
        A383 & $0.188$ & $4.1 \pm 0.1$ & $3.11^{+0.88}_{-0.69}$ & $5.59^{+6.31}_{-3.51}$ \\
        A520 & $0.203$ & $7.9 \pm 0.2$ & $4.20^{+2.30}_{-1.70}$ & $2.20^{+1.10}_{-0.80}$ \\
        A521 & $0.248$ & $6.1 \pm 0.1$ & $4.58^{+1.00}_{-0.88}$ & $2.18^{+1.13}_{-0.83}$ \\
        A586 & $0.171$ & $7.8^{+1.0}_{-0.8}$ & $6.29^{+2.26}_{-1.69}$ & $4.90^{+5.82}_{-3.16}$ \\
        A611 & $0.288$ & $7.1^{+0.6}_{-0.5}$ & $5.47^{+1.31}_{-1.11}$ & $3.01^{+1.99}_{-1.37}$ \\
        A697 & $0.282$ & $8.8^{+0.7}_{-0.6}$ & $9.73^{+1.86}_{-1.61}$ & $2.17^{+0.94}_{-0.73}$ \\
        A963 & $0.206$ & $5.6 \pm 0.1$ & $5.10^{+1.70}_{-1.40}$ & $2.10 \pm 0.40$ \\
        A1835 & $0.253$ & $8.4 \pm 0.1$ & $10.86^{+2.53}_{-2.08}$ & $2.35^{+1.27}_{-0.93}$ \\
        A1914 & $0.171$ & $8.5 \pm 0.2$ & $4.20^{+2.90}_{-2.00}$ & $3.20^{+0.70}_{-0.60}$ \\
        A2034 & $0.113$ & $6.4 \pm 0.2$ & $4.30^{+4.00}_{-2.40}$ & $2.90^{+0.90}_{-0.70}$ \\
        A2219 & $0.228$ & $9.6^{+0.3}_{-0.2}$ & $7.75^{+1.89}_{-1.60}$ & $4.59^{+4.11}_{-2.53}$ \\
        A2261 & $0.224$ & $6.1^{+0.6}_{-0.5}$ & $7.97^{+1.51}_{-1.31}$ & $4.31^{+2.37}_{-1.72}$ \\
        A2390 & $0.231$ & $9.1 \pm 0.2$ & $6.92^{+1.50}_{-1.29}$ & $4.40^{+2.63}_{-1.86}$ \\
        A2631 & $0.278$ & $7.5^{+0.4}_{-0.2}$ & $4.54^{+0.89}_{-0.78}$ & $5.57^{+4.24}_{-2.78}$ \\
        RX J1720 & $0.164$ & $5.9 \pm 0.1$ & $3.48^{+1.28}_{-0.99}$ & $4.87^{+6.98}_{-3.45}$ \\
        RX J2129 & $0.235$ & $5.6 \pm 0.1$ & $5.29^{+1.76}_{-1.38}$ & $4.67^{+6.21}_{-3.19}$ \\
        ZwCl 1454 & $0.258$ & $4.6 \pm 0.1$ & $2.80^{+1.39}_{-1.03}$ & $2.18^{+3.25}_{-1.57}$ \\
        ZwCl 1459 & $0.290$ & $6.4 \pm 0.2$ & $3.77^{+1.17}_{-0.98}$ & $4.10^{+4.56}_{-2.56}$ \\        
        \hline 
    \end{tabular}

    \medskip
    X-ray data from \citet{Ade:2012vda,Landry2013,Kenneth2008}. 
\end{table}
We assume flat distributions within the range $[x-\Delta x_2, x+\Delta x_1]$ for the inputs in the form $x^{+\Delta x_1}_{-\Delta x_2}$ in the generation of random realisations for our Monte Carlo method.

Inspection of the final distributions of virial ratios and interaction strengths suggests the use of log-normal distributions to fit (the negative of) the data. However, due to the nature of these quantities and their domains, we include a location parameter to allow the distribution to be shifted from the origin. The log-normal PDF is then
\begin{align}
    f_X(x; \mu,\sigma,\loc) = \frac{e^{- \left[\ln(x-\loc)-\mu\right]^2/2\sigma^2}}{(x-\loc)\sigma\sqrt{2\pi}}
\end{align}
with $\loc$ being the location parameter. We take the 68\% most probable values and the maximum probability of this log-normal PDF to yield the resulting value $x^{+\Delta x_1}_{-\Delta x_2}$ of the quantity $X$. The fits obtained are especially good for the interaction strengths, as we show in Sec.~\ref{sec:interaction}.

In addition, we introduce two selection criteria which we apply to the values of the interaction strength obtained with this method to conserve realisations: one physical, discussed in Sec.~\ref{sss:intmodel}, selects $\xi \le 3$, while the other avoids numerical problems, discussed in Sec.~\ref{sss:nvmodel}, by keeping only $\xi \ge -200$ (see discussion in Sec.~\ref{sec:interaction}).

\subsubsection{Reliability of the results}
Our analysis considers samples of 2600 random realisations of each cluster, but some of them happen to have no solution for $\xi$, or to have a solution outside the domain established by Eq.~(\ref{eq:L-I}). These realisations are removed from the analysis, leading to considerably smaller samples for some clusters. That is the case for A68, A115, A520 and A1914, for which we are left with only about 200 realisations. A possible explanation for such a large fraction of these samples not having a physical solution for $\xi$ could be linked with the dynamical activity of those clusters \citep[e.g.][]{Mark:2005,Barrena:2013}, so their virial states are not as close to equilibrium for our method to be applicable.

The gas distribution in clusters can be used as a probe of the recent past 
dynamical activity of a cluster, since the gas responds quickly to 
perturbations in the gravitational potential, for instance, due to cluster 
merger and/or collision \citep{AndradeSantos2012}. Visual inspections of 
Chandra X-ray images show that all clusters except A115, A520 and A1914 have rather 
undisturbed and symmetric gas distribution, suggesting that they are not 
dynamically active. 

For comparison purposes, we define a ``success rate'' (SR) as the fraction of realisations satisfying our selection criteria in the total generated for each cluster. Clusters like A1835, A209 and A2261 present this fraction very close or equal to $1$. The success rates for all clusters are presented in Table~\ref{tab:successrate}. In the histograms, we use a number of bins proportional to the size of the samples.
\begin{table*}
    \begin{minipage}{150mm}
        \caption{Fraction of Monte Carlo-produced realisations for each cluster.}
        \label{tab:successrate}
        \begin{tabular}{@{}lccccccccccc@{}}
            \hline
            Cluster  & A115 & A1835 & A1914 & A2034 & A209 & A2219 & A2261 & A2390 & A2631 & A267 & A383 \\
            SR & $0.06$ & $1.00$ & $0.08$ & $0.58$ & $1.00$ & $0.80$ & $1.00$ & $0.74$ & $0.68$ & $0.60$ & $1.00$ \\
            \hline
            Cluster & A520 & A521 & A586 & A611 & A68 & A697 & A963 & RX J1720 & RX J2129 & ZwCl 1454 & ZwCl 1459 \\ 
            SR & $0.05$ & $0.67$ & $0.85$ & $0.80$ & $0.11$ & $0.98$ & $0.91$ & $0.72$ & $1.00$ & $0.64$ & $0.63$ \\             
            \hline 
        \end{tabular}

        \medskip
        The SR here presented were computed for our Monte Carlo samples of size $2600$. Tests have shown that there is no significant variation of the SR with the size of the sample.
    \end{minipage}
\end{table*}

In order to evaluate the consistency of our method, we consider a cluster from $N$-body simulation, similar to those of \citet{Machado:2013}, in a cosmology with $\Omega_{M} = 0.3$, $\Omega_{DE} = 0.7$, $h = 0.72$ and no interaction in the dark sector, so the virial ratio should be very close\footnote{Some variations can be introduced by the effects of projection translating from simulation to observables.} to $-0.5$ and interaction compatible with zero. The data for this cluster are $M_{200} = 18.0 ~h^{-1}10^{14}M_{\odot}$, $z = 0.0$, $c = 3.0$ and $T_X = 7.3 \pm 0.8~\rmn{keV}$. The uncertainty in the temperature comes from the $\sigma_X$--$T_X$ scatter relation \citep{Xue:2002kd}
\begin{align}
    \sigma_X = 10^{2.49 \pm 0.02} T_X^{0.65 \pm 0.03},
\end{align}
from which $T_X$ was computed for a one-dimensional velocity dispersion of $\sigma_X = 1125~\rmn{km/s}$.
Because the observed virial ratio is linear with the temperature, the only source of errors in this case, its histogram for all random realisations produced in our code reflects clearly the uniform distribution assigned to the input temperature. Fitting that uniform distribution we find a virial ratio of $-0.47 \pm 0.04$ from the central 68\% most probable values.

With the analysis of Sec.~\ref{sss:generalfit}, for the interaction strength we get $\xi = -0.05^{+0.54}_{-0.69}$, therefore compatible with the simulation. The theoretical virial ratio is also in accordance with the classic value, $\left( \frac{\rho_K}{\rho_W}\right)_{th} = -0.51 \pm 0.05$, while the DfE being $-0.08 \pm 0.06$ satisfies our condition (\ref{eq:Close2EqCond}). 

\section{Analysis of the Results}
\label{sec:Analysis-of-the}

In what follows we discuss the outcome of our analysis starting with the observed virial ratios (hereafter OVR), their theoretical estimates (which we denote by TVR) from combining their DfE factors and their interaction strengths $\xi$.
Throughout this section, we present the constraints on the virial ratios, interaction and departure from equilibrium in Figs.~\ref{fig:MCVR_all}, \ref{fig:xi}, \ref{fig:vrout} and \ref{fig:OB}, all obtained for each cluster according to the method described in the previous section. These results are summarised in Table~\ref{tab:results}. The detailed distributions of some of these quantities are shown in the Appendix~\ref{app:hist}.
\begin{table}
    \caption{Virial ratios, interactions, theoretical virial ratios and departure from equilibrium from the log-normal fits.}
    \label{tab:results}
    \tabcolsep 3pt
    \begin{tabular}{@{}lcccc@{}}
        \hline
        Cluster &   OVR &   $\xi$  &  TVR & DfE  \\
        \hline
        A520 & $-0.96 \pm 0.02$ & $-53.28^{+19.87}_{-44.20}$ & $-0.96 \pm 0.02$ & $-0.003 \pm 0.003$ \\ 
        A1914 & $-0.95 \pm 0.02$ & $-48.60^{+18.59}_{-43.09}$ & $-0.96 \pm 0.02$ & $-0.004 \pm 0.003$ \\ 
        A115 & $-0.94 \pm 0.03$ & $-35.72^{+17.85}_{-43.34}$ & $-0.95 \pm 0.03$ & $-0.01 \pm 0.01$ \\ 
        A68 & $-0.93 \pm 0.03$ & $-31.62^{+13.90}_{-37.11}$ & $-0.94 \pm 0.03$ & $-0.01 \pm 0.01$ \\ 
        A521 & $-0.89 \pm 0.05$ & $-13.14^{+5.67}_{-24.80}$ & $-0.90 \pm 0.05$ & $-0.01 \pm 0.03$ \\ 
        A267 & $-0.88 \pm 0.06$ & $-10.50^{+5.45}_{-24.20}$ & $-0.89 \pm 0.06$ & $-0.01 \pm 0.04$ \\ 
        A2390 & $-0.87 \pm 0.06$ & $-13.08^{+5.90}_{-22.60}$ & $-0.89 \pm 0.05$ & $-0.05 \pm 0.04$ \\ 
        A2631 & $-0.87 \pm 0.07$ & $-11.56^{+5.75}_{-22.91}$ & $-0.89 \pm 0.06$ & $-0.04 \pm 0.05$ \\ 
        A611 & $-0.85^{+0.07}_{-0.08}$ & $-8.76^{+4.58}_{-20.08}$ & $-0.87 \pm 0.07$ & $-0.03 \pm 0.05$ \\ 
        ZwCl 1459 & $-0.85 \pm 0.08$ & $-7.41^{+4.43}_{-21.56}$ & $-0.86^{+0.07}_{-0.08}$ & $-0.02 \pm 0.06$ \\ 
        A2219 & $-0.84 \pm 0.08$ & $-9.28^{+4.48}_{-18.55}$ & $-0.87 \pm 0.06$ & $-0.07 \pm 0.05$ \\ 
        ZwCl 1454 & $-0.82 \pm 0.09$ & $-3.95^{+2.42}_{-15.49}$ & $-0.82^{+0.09}_{-0.10}$ & $0.02 \pm 0.06$ \\ 
        RX J1720 & $-0.81 \pm 0.10$ & $-4.47^{+2.94}_{-16.34}$ & $-0.83 \pm 0.09$ & $-0.02 \pm 0.07$ \\ 
        A697 & $-0.81 \pm 0.07$ & $-8.17^{+3.62}_{-10.75}$ & $-0.84 \pm 0.06$ & $-0.04 \pm 0.06$ \\ 
        A963 & $-0.78^{+0.07}_{-0.10}$ & $-4.59^{+1.98}_{-13.39}$ & $-0.80^{+0.08}_{-0.09}$ & $-0.01 \pm 0.05$ \\ 
        A2034 & $-0.77^{+0.08}_{-0.10}$ & $-4.37^{+2.23}_{-13.35}$ & $-0.80^{+0.08}_{-0.09}$ & $-0.02 \pm 0.06$ \\ 
        A586 & $-0.77 \pm 0.11$ & $-3.83^{+2.74}_{-12.51}$ & $-0.80^{+0.09}_{-0.10}$ & $-0.07 \pm 0.08$ \\ 
        A1835 & $-0.70^{+0.06}_{-0.07}$ & $-4.84^{+2.10}_{-3.78}$ & $-0.75 \pm 0.06$ & $-0.07 \pm 0.06$ \\ 
        A383 & $-0.61^{+0.08}_{-0.11}$ & $-0.41^{+1.04}_{-3.44}$ & $-0.60^{+0.09}_{-0.12}$ & $0.05 \pm 0.10$ \\ 
        A209 & $-0.60 \pm 0.05$ & $-1.92^{+1.01}_{-1.41}$ & $-0.64 \pm 0.05$ & $-0.05 \pm 0.08$ \\ 
        RX J2129 & $-0.60^{+0.09}_{-0.12}$ & $-0.68^{+1.23}_{-4.04}$ & $-0.63^{+0.10}_{-0.12}$ & $-0.02 \pm 0.10$ \\ 
        A2261 & $-0.56^{+0.06}_{-0.07}$ & $-0.99^{+1.00}_{-1.55}$ & $-0.60 \pm 0.07$ & $-0.03 \pm 0.09$ \\ 
        TOTAL & --- & $-1.99^{+2.56}_{-16.00}$ & $-0.79 \pm 0.13$ & --- \\  
        \hline
    \end{tabular}
\end{table}

\subsection{The observed virial ratios}
\label{ss:MCVR}

\begin{figure}
    \centering
    \subfigure[\label{fig:MCVR_A2261}OVR for the cluster A2261 from a sample of 2600 random realisations. $\frac{\rho_K}{\rho_W} = -0.56^{+0.06}_{-0.07}$. ]{\includegraphics[width=\columnwidth]{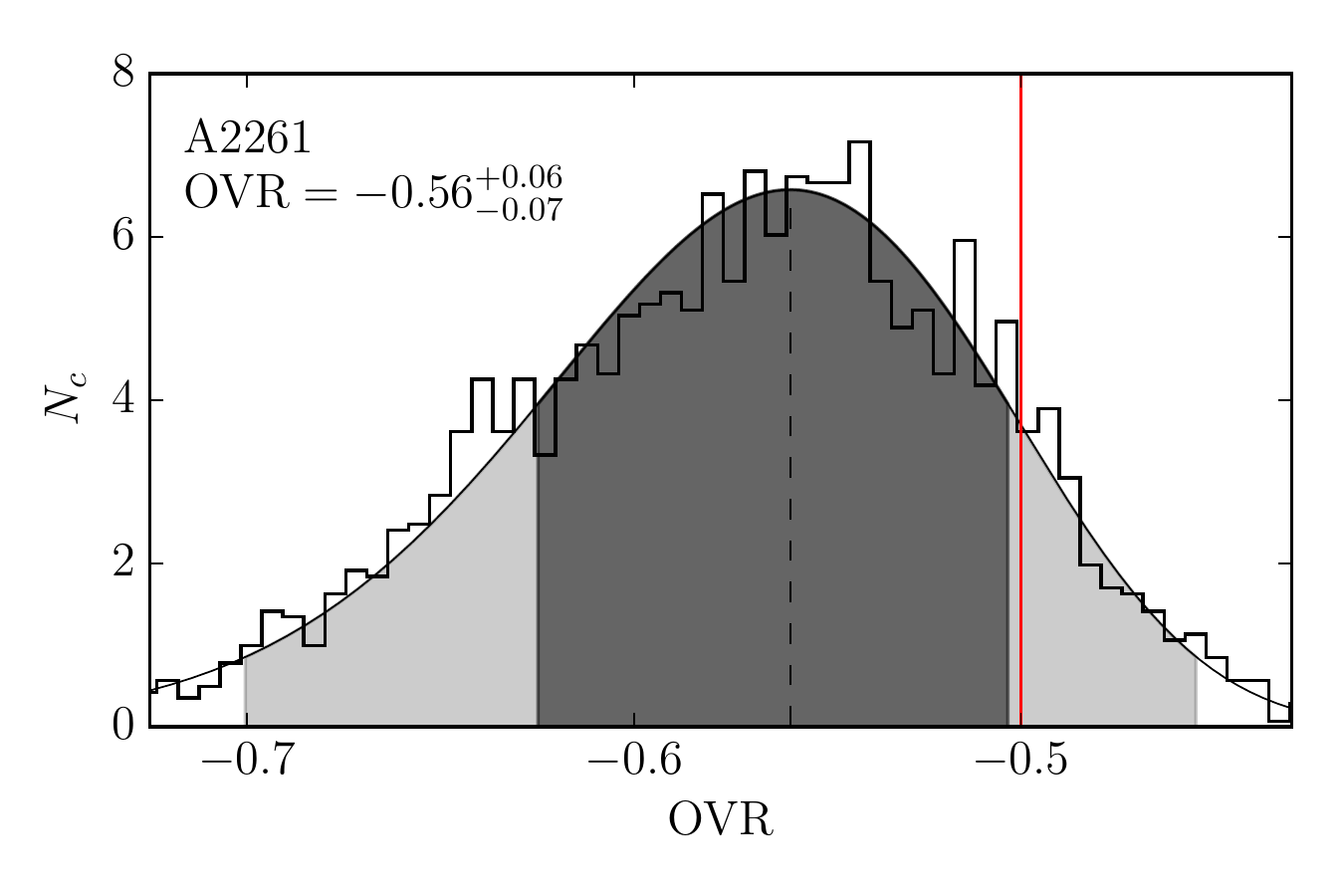}} \\
    \subfigure[\label{fig:MCVR_all}OVR with error bars indicating 95\% and 68\% C.L.~for all clusters.]{\includegraphics[width=\columnwidth]{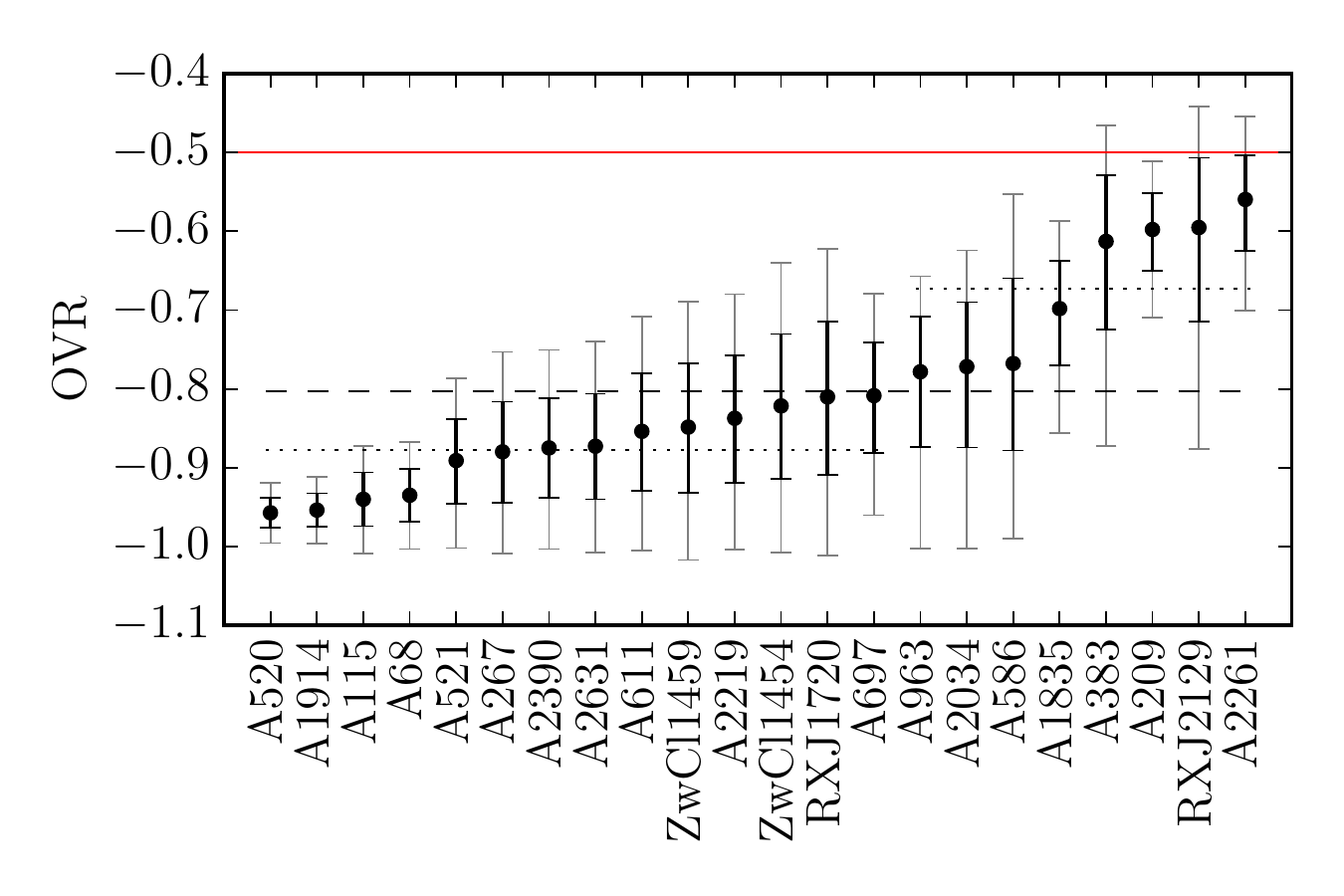}}
    \caption{\subref{fig:MCVR_A2261} shows in detail the distribution of the OVR for the cluster A2261 when we apply the Monte Carlo method (Sec.~\ref{ss:MonteCarlo2}). $N_c$ is the normalised count of Monte Carlo clusters per bin of OVR. The dark and light shaded areas correspond to 68\% and 95\% C.L. The dashed vertical line indicates the most probable value while the red solid line denotes the classic value. \subref{fig:MCVR_all}  presents the most probable values and confidence levels for each cluster. The grey and black error bars give, respectively, the 95\% and 68\% C.L. The mean of the most probable values is signaled by the dashed black line. The two dotted lines show the means for the two groups of clusters defined in the text and the solid red line marks the classic value.}
    \label{fig:MCVR}
\end{figure}

The OVR is obtained from applying the method described in Sec.~\ref{ss:MonteCarlo2} to Eq.~(\ref{eq:Virialratio}). 
The histograms of OVR produced for each cluster are very similar to what we obtain for their theoretical counterparts TVR (see Appendix~\ref{app:hist}).
As an example we present in Fig.~\ref{fig:MCVR_A2261} the distribution obtained for the cluster A2261.
It represents the histogram distribution of the OVR obtained from our Monte Carlo sampling of mass, temperature and concentration ranges. Superimposed is the log-normal fit, with light-shaded area corresponding to 95\% C.L., and dark-shaded area emphasising 68\% C.L. The red vertical line marks the theoretical non-interacting value, while the dashed line gives the most likely value.

We summarise the results of the OVR with their corresponding asymmetrical errors in Fig.~\ref{fig:MCVR_all}, where we have shown the theoretical non-interacting virial ratio as a horizontal red line.
We have ordered the clusters by increasing OVR and keep this order for the rest of the work.

We have represented the mean value for the whole sample with a dashed line.
We identify two groups of similar virial ratios separated by the global mean and for each group we represent their means by the dotted lines. The dispersion of the ratios may reflect the diversity of the equilibrium conditions. However, the first group seems to have less scatter than the second one. 

With this robust non-linear treatment of error propagation, all clusters
exclude $-0.5$ at $1\sigma$.

\subsection{The interaction strength}
\label{sec:interaction}
As previously mentioned we solve for $\xi$ Eq.~(\ref{eq:Virial+Out}) with the departure from equilibrium term (\ref{eq:Out(xi)}) using the Monte Carlo method of Sec.~\ref{ss:MonteCarlo2}.
Eq.~(\ref{eq:Coupling}) is singular at $\frac{\rho_K}{\rho_W} = - 1$, which corresponds to $\xi$ infinite. This is a limitation of our method that we deal with by restricting the interaction strength to $|\xi| \le 200$ such that the histograms would be legible. Note that this introduces a cut in the histograms of the theoretical virial ratios TVR (Appendix~\ref{app:hist}), which reflects the limitation of our method.

\begin{figure}
    \centering
    \includegraphics[width=\columnwidth]{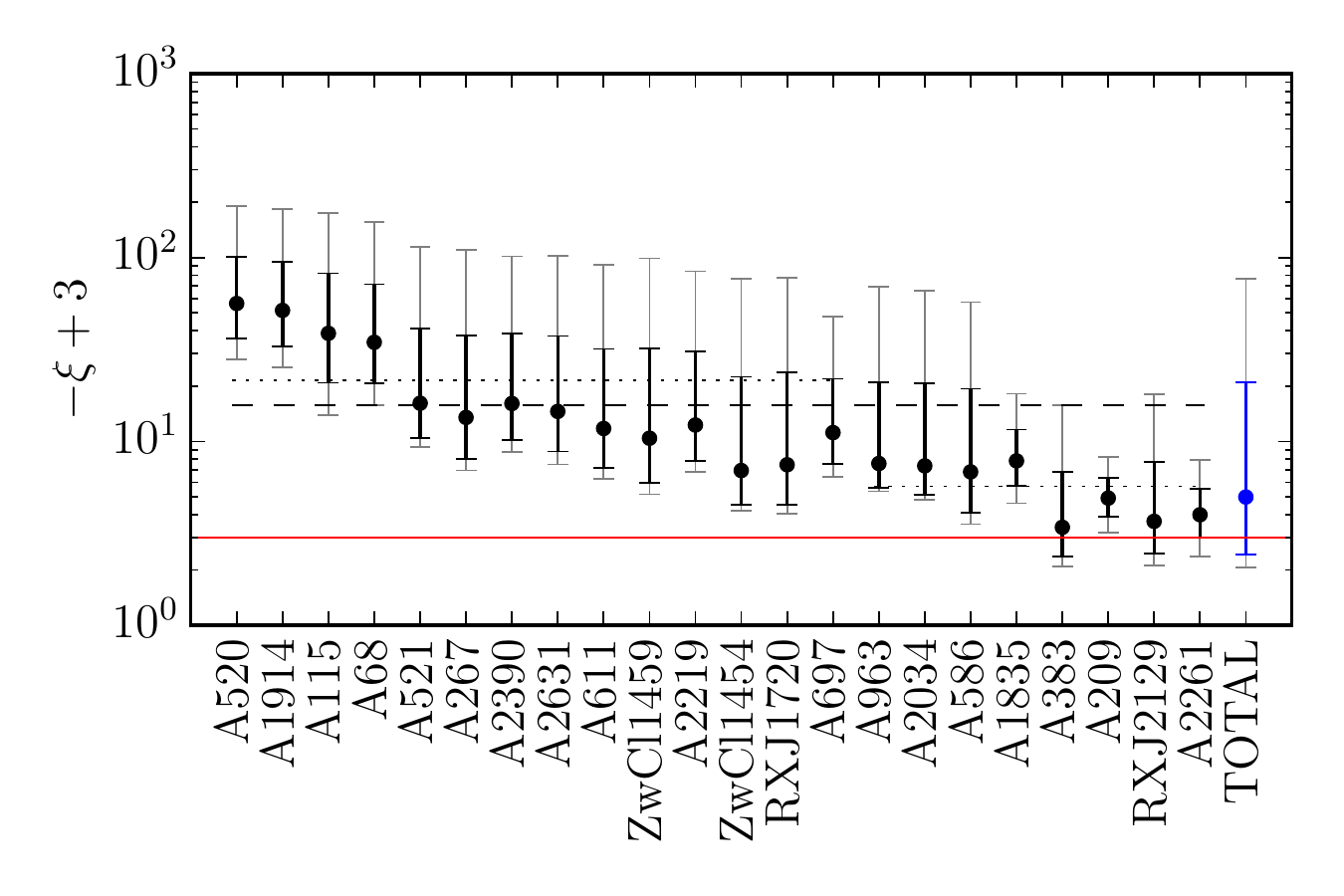}
    \caption{Interaction strengths with error bars for each cluster in our sample. The grey and black error bars give, respectively, the 95\% and 68\% C.L. The most probable value and error bars of the compounded distribution are shown at the right in blue. Again, dotted horizontal lines represent the means for each group of clusters, while the dashed line marks the overall mean. The solid red line marks $\xi=0$. We plot $-\xi+3$ rather than $\xi$ to enable visualization in log scale.}
    \label{fig:xi}
\end{figure}

In Fig.~\ref{fig:xi} we plot the most probable value of $-\xi+3$ for all clusters to keep the log scale, together with the 68\% and 95\% C.L.~corresponding to, respectively, black and grey error bars. The red solid line indicates the $\xi=0$ absence of interaction. The mean value for the whole sample is given by the dashed line.
Each group previously singled out also present their group mean as dotted lines.
Finally, we add the most probable value and error bars of the compounded distribution in blue.

Of the 22 clusters, all of them except A383, RX J2129 and A2261 (marginally) display a $1\sigma$ detection favouring negative $\xi$.

In this model, the interaction strength should be the same for all clusters. The global mean is compatible with 13 of the 22 clusters: A521, A267, A2390, A2631, A611, ZwCl 1459, A2219, ZwCl 1454, RX J1720, A697, A963, A2034, and A586. It gives $-12.78$. However, three of the clusters, as well as the compounded distribution that yields a value for $\xi = -1.99^{+2.56}_{-16.00}$, displays compatibility with no interaction.
This points to problems in our method, namely when it assumed small deviation from equilibrium while the results have important variations.
    Actually, the lowest boundary is limited by the results of \citet{Salvatelli2013} to\footnote{\citeauthor{Salvatelli2013} have found $-2.70 < \xi < -0.66$ at 95\% C.L.} $-2.70$ \citep[see also][]{AndreElcio2014PRD89}.
Nevertheless, we concentrate on the present scheme and leave the solutions to a forthcoming work.

\subsection{The theoretical virial ratios}
\label{ss:theoreticalVR}
\begin{figure}
    \centering
    \includegraphics[width=\columnwidth]{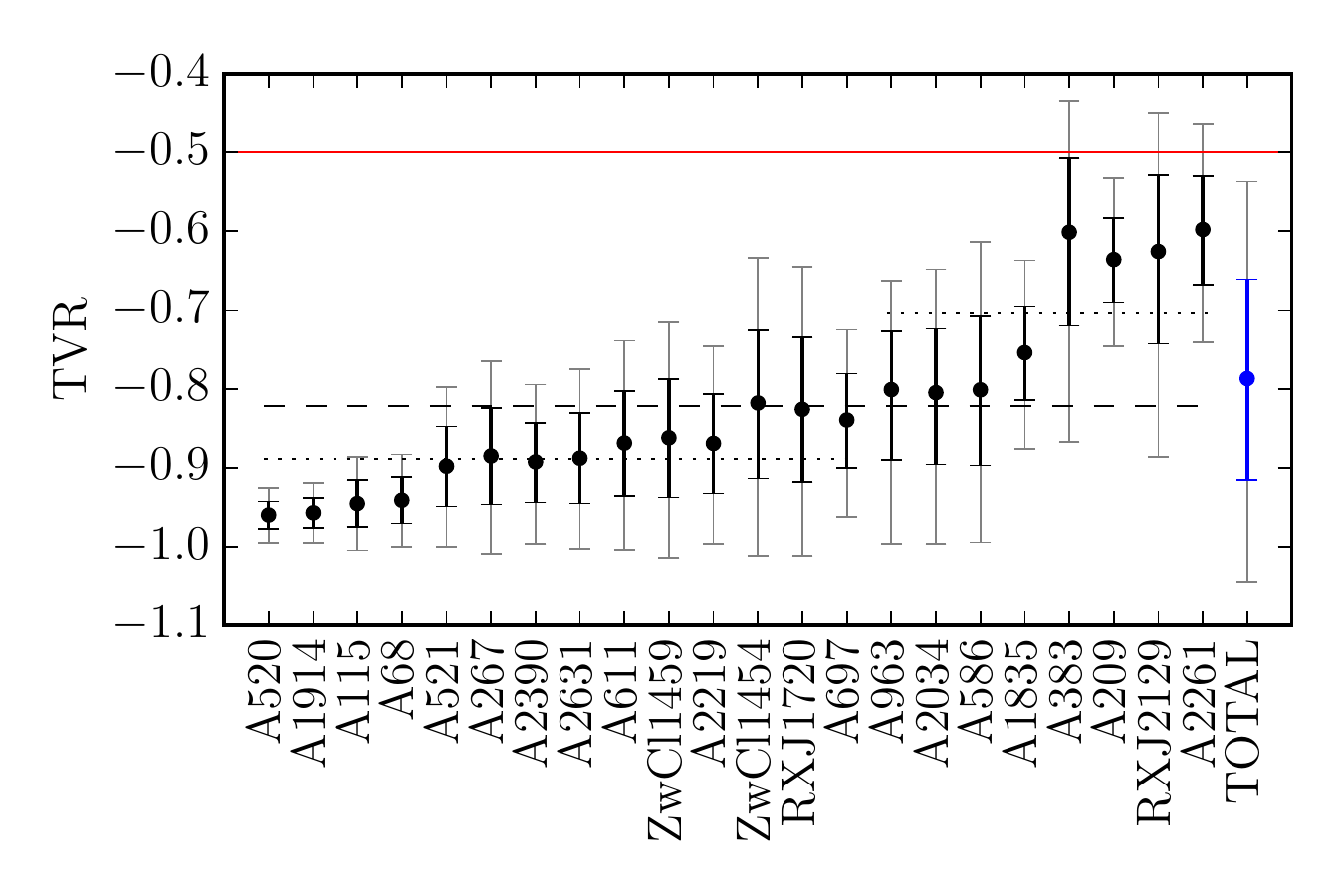}
    \caption{TVR with confidence intervals for each cluster in our sample and for the compounded distribution. Grey and black error bars give, respectively, 95\% and 68\% C.L. The most probable value and error bars of the compounded distribution are shown at the right in blue. Dotted horizontal lines represent means for each group, while dashed line marks overall mean. The solid red line marks the non-interaction case.}
    \label{fig:vrout}
\end{figure}

Armed with the results from the previous section, we compute with Eq.~(\ref{eq:thVR}) the TVR that each cluster would have at perfect equilibrium in the presence of interaction. 
Their corresponding distributions are shown in the Appendix~\ref{app:hist}. 
As those histogram are very similar to the observed ones (Sec.~\ref{ss:MCVR}), the following comments can be applied to both. The reasons for these similarities are discussed in Sec.~\ref{subsec:outofbalance}. 
We plot in Fig.~\ref{fig:vrout} the most probable values of TVR for all clusters, together with the 68\% and 95\% C.L.~corresponding to, respectively, black and grey error bars.
The red solid line indicates the $-0.5$ classic virial value, in absence of interaction. The mean value for the whole sample is given by the dashed line. In the same way as with the interaction strengths, we compounded the distributions of the TVR for all the clusters, as they should all be equal to a universal value, and show the most probable value and its error bars in blue, at the right of the panel.

At this level, all the clusters exclude $-0.5$ at $1\sigma$, which confirms the result from Sec.~\ref{ss:MCVR}.
However, 16 out of the 22 clusters present log-normal fits which reflects poorly the underlying distributions:
A520, A1914, A115, A68, A521, A267, A2390, A2631, A611, ZwCl 1459, A2219, ZwCl 1454, RX J1720, A963, A2034 and A586 (see Appendix~\ref{app:hist}).
For all of those problematic distributions the log-normal fits break down for virial ratios in the proximity of $-1$. This is related to the singularity in Eq.~(\ref{eq:Coupling}).
In addition, the compounded TVR points towards a single value of $-0.79 \pm 0.13$, which represents a detection at $2\sigma$, in contradiction with the results of the previous section.
All this suggests a problem with our method that assumed small deviation from equilibrium, as previously pointed out.

In the following section, we discuss how the observed and the theoretical virial ratios are so similar.

\subsection{The departure from equilibrium factors}
\label{subsec:outofbalance}
\begin{figure}
    \centering
    \includegraphics[width=\columnwidth]{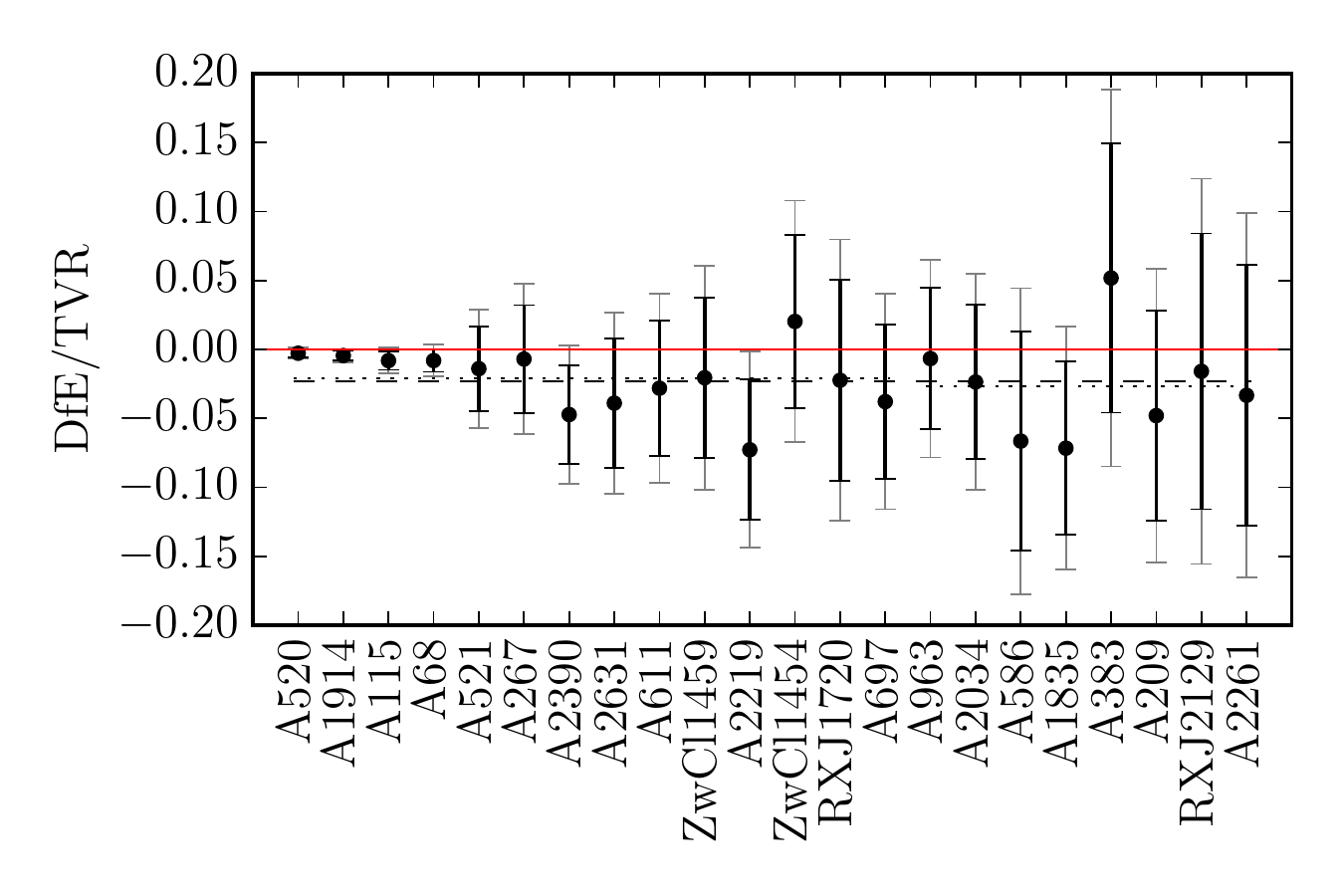}
    \caption{DfE factors relative to the TVR with confidence intervals for each cluster in our sample. Grey and black error bars give, respectively, 95\% and 68\% C.L. Dotted horizontal lines represent means for each group, while dashed line marks overall mean. The solid red line marks the absence of deviation.}
    \label{fig:OB}
\end{figure}

Eq.~(\ref{eq:Out(xi)}) with the results of Sec.~\ref{sec:interaction} allows us to compute the DfE factor for each cluster.
The values of this factor relative to their theoretical virial ratio,
\begin{align}
    \frac{\text{DfE}}{\text{TVR}} = \frac{-\left(2-\frac {\xi} {3}\right)^{-2}  \dot\rho_W / H\rho_W}{\bigl(\rho_K / \rho_W\bigr)_{th}} = \frac{\dot \rho_W/ H \rho_W}{(1-\xi/3)(2-\xi/3)},
\end{align}
are presented in Fig.~\ref{fig:OB}. 
Except for A1914, A115, A2390, A2219, and A1835, all those relative departures appear compatible with zero.
We should note as well that ZwCl 1454 and A383 are the only clusters with positive DfE.
For this figure and for numerical reasons, we fit the distributions for each cluster with uniform distributions so as to evaluate the order of magnitude of those departures and produce the values displayed on Fig.~\ref{fig:OB}.
Although not good fits, these uniform distributions enable us to show how small those values are, validating our hypothesis (\ref{eq:Close2EqCond}).
This explains the similarities between OVR and TVR, as seen in comparing Fig.~\ref{fig:MCVR_all} and Fig.~\ref{fig:vrout}.

\section{Discussion and Conclusions}
\label{sec:Discussion-and-Conclusions}

We analysed the virial ratios of a set of clusters using a simple model based on the Layzer--Irvine equation 
\citep{Bertolami:2007zm,LeDelliou:2007am,Bertolami:2008rz,Bertolami:2007tq,Abdalla:2007rd,Abdalla:2009mt,He:2010ta,Bertolami:2012yp}, using weak-lensing mass profiles and intracluster gas temperatures from optical and X-ray observations \citep{Okabe25062010,Ade:2012vda}. Our treatment involved assessing the virial balance of each cluster as well as their equilibrium state, using a Monte Carlo statistical analysis on the data.
Note that the equilibrium state evaluation should encompass every single known and unknown source of deviation from balance.

Our method, a first proof of concept for out of equilibrium virial evaluation, enabled us to find mild evidence for an interacting dark sector in the virial balance of those clusters, however yielding only small amplitudes of departure from equilibrium: although the compounded distribution of all clusters would accommodate $\xi = 0$, a majority of the individual clusters, of their virial ratios and of the compounded evaluation of the virial ratio all point towards a negative interaction. The compounded estimates give us $\xi = -1.99^{+2.56}_{-16.00}$, which is not a detection, but TVR $\bigl(\rho_K/\rho_W\bigr)_{th} =-0.79\pm 0.13$, which is a detection at $2\sigma$.
This tension between the compounded results for the interaction strength and the theoretical virial ratio, while the latter is constructed out of the former, points to the main problem in our results: despite the scatter in the values of virial ratios, the departure from equilibrium factors remain small, as imposed in our hypotheses.
In addition, our method contains an unphysical singularity at $\frac{\rho_K}{\rho_W}=-1$ in Eq.~(\ref{eq:Coupling}). These problems, in spite of encouraging results, call for follow up work which should remove the small departure from equilibrium hypothesis, as well as the singularity we introduced in this work for $\rho_K =-\rho_W$.

Most observed clusters appear to be somewhat perturbed systems and are maybe still forming (accreting mass), which is expected in the current standard cosmological scenario.
Our approach, that treat clusters as out of equilibrium systems, is therefore natural, despite the measurement of such departure not following the observed wide variations in virial states. The tension between the results of our measured departure from the classic virial ratio and our measured interaction strength reflects both that our method shows potential but also has room for improvement and we can expect that accommodation for large departures should enable the use of much larger samples that can enhance statistically the significance of the results.

\section{Acknowledgements}

The work of M.Le~D.~has been supported by
FAPESP (2011/24089-5) and PNPD/CAPES20132029. M.Le~D.~also wishes to acknowledge DFMA/IF/USP and IFT/UNESP.
R.J.F.M.~and G.B.L.N.~thank CAPES for support. G.B.L.N.~also wish to thank CNPq. E.A.~acknowledges FAPESP.
Finally, we thank Rubens Machado for his simulation.

\appendix

\section{Histograms of the virial ratios and interaction strengths}
\label{app:hist}
\begin{figure*}
    \centering
    \includegraphics[width=0.9\textwidth]{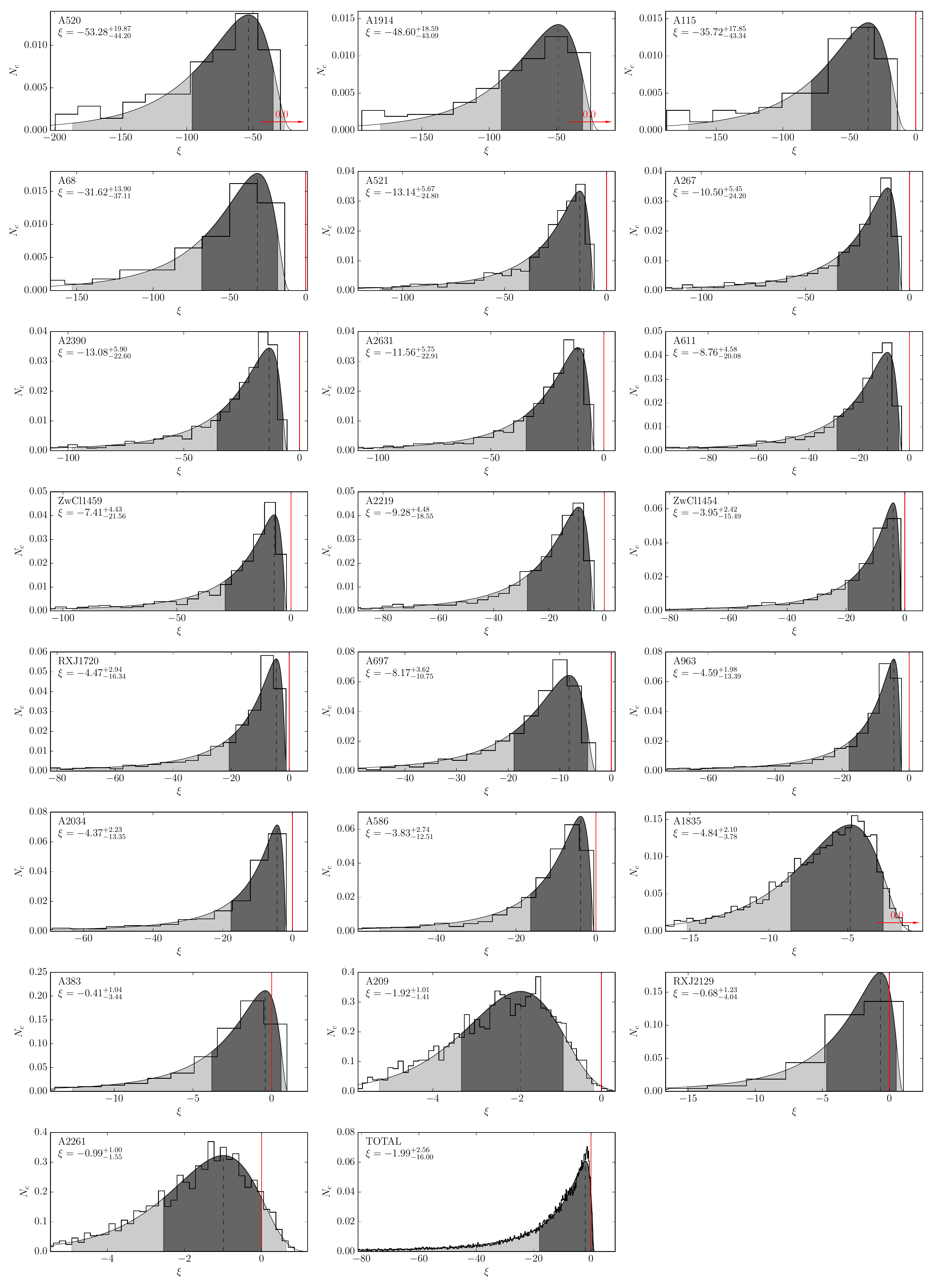}
    \caption{Histograms for the distributions of interaction strength $\xi$ with their log-normal fits for each cluster. $N_c$ is the normalised count of Monte Carlo clusters per bin of $\xi$. The lowest rightmost panel presents the cumulated histogram of all the clusters. The dark and light shaded areas mark the 68\% and 95\% C.L. The dashed lines point the most probable values, indicated inside the frames with errors referring to the 68\% C.L., while the solid red line (red arrow) marks (points to) the $\xi=0$ position.}
    \label{fig:allxi6895}
\end{figure*}

\begin{figure*}
    \centering
    \includegraphics[width=0.9\textwidth]{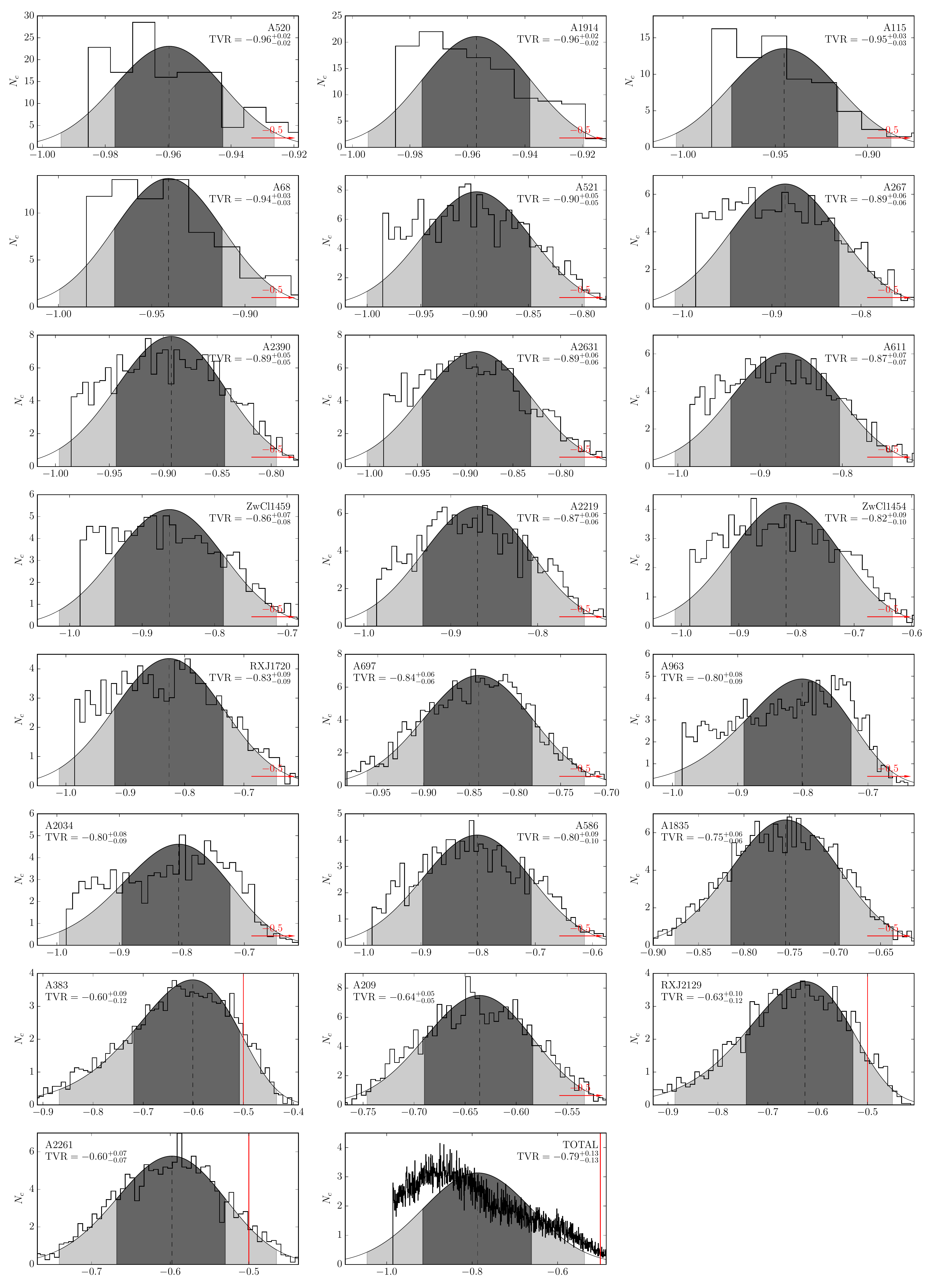}
    \caption{Histograms for the distributions of the TVR with their log-normal fits for each cluster. $N_c$ is the normalised count of Monte Carlo clusters per bin of TVR. The lowest rightmost panel presents the cumulated histogram of all the clusters. The dark and light shaded areas mark the 68\% and 95\% C.L. The dashed lines point the most probable values, indicated inside the frames with the errors referring to the 68\% C.L., while the solid red line (red arrow) marks (points to) the position of the no interaction classic virial ratio.}
    \label{fig:allvrout6895}
\end{figure*}

In Figs.~\ref{fig:allxi6895} and \ref{fig:allvrout6895} we show, respectively, the histograms of interaction strengths and theoretical virial ratios for all clusters, their lowest rightmost panels displaying the compounded distribution for all the studied clusters. We keep conventions of light and dark shaded areas representing 95\% and 68\% C.L., the dashed line for the most probable value and the red solid line (when visible) to indicate absence of interaction.

\label{lastpage}
\end{document}